# GraphFuzz: Automated Testing of Graph Algorithm Implementations with Differential Fuzzing and Lightweight Feedback

WENQI YAN, The University of Melbourne, Australia
MANUEL RIGGER, National University of Singapore, Singapore
ANTHONY WIRTH, The University of Sydney, Australia
VAN-THUAN PHAM, The University of Melbourne, Australia

Graph algorithms, such as shortest path finding, play a crucial role in enabling essential applications and services like infrastructure planning and navigation, making their correctness important. However, thoroughly testing graph algorithm implementations poses several challenges, including their vast input space (i.e., arbitrary graphs). Moreover, through our preliminary study, we find that just a few automatically generated graphs (less than 10) could be enough to cover the code of many graph algorithm implementations, rendering the code coverage-guided fuzzing approach—one of the state-of-the-art search algorithms—less efficient than expected.

To tackle these challenges, we introduce *GraphFuzz*, the first automated feedback-guided fuzzing framework for graph algorithm implementations. Our key innovation lies in identifying lightweight and algorithm-specific feedback signals to combine with or completely replace the code coverage feedback to enhance the diversity of the test corpus, thereby speeding up the bug-finding process. This novel idea also allows *GraphFuzz* to effectively work in both black-box (i.e., no code coverage instrumentation/collection is required) and grey-box setups. *GraphFuzz* uses differential testing to detect logic bugs, in addition to crash-triggering bugs captured through exceptions. Our evaluation demonstrates the effectiveness of *GraphFuzz*. The tool has successfully discovered 12 previously unknown bugs, including 6 logic bugs, in 9 graph algorithm implementations in two popular graph libraries, NetworkX and iGraph. All of them have been confirmed and and 11 bugs have been rectified by the libraries' maintainers.

## 1 Introduction

The correctness of graph algorithm implementations is crucial for many essential services [25]. These services range from mapping the links of the Internet and designing infrastructure networks to enabling navigation tools and optimizing logistical processes in business operations, among many

Authors' Contact Information: Wenqi Yan, wenqi.yan@student.unimelb.edu.au, The University of Melbourne, Australia; Manuel Rigger, rigger@nus.edu.sg, National University of Singapore, Singapore; Anthony Wirth, anthony.wirth@sydney. edu.au, The University of Sydney, Australia; Van-Thuan Pham, thuan.pham@unimelb.edu.au, The University of Melbourne, Australia.

other critical applications. However, to the best of our knowledge, manual test design is the sole testing method employed thus far to validate graph implementations[1]. It involves crafting specific graphs, computing expected results, and then validating whether the output of the algorithm implementation matches the pre-computed value. This approach is labor-intensive, error-prone as the human-specified expected results might be wrong, and does not scale due to the following challenges.

> *"This morning I was talking to a guy in Microsoft Research who is pretty famous. He described a neat project he worked on that essentially boiled down to a very complex graph shortest-path problem. That got my mind churning on the general problem of graph algorithms and testing graph algorithm implementations.[...] I hadn't looked at a graph problem in a while so I decided to code up the more-or-less canonical example: Dijkstra's algorithm [...]. Even with a tiny graph it's easy for a human to make a mistake. So, what about testing graph algorithms? In some ways this is a classic problem: a large number of inputs, possibility of bad arguments, many assumptions, and so on."*
>
> – J. D. McCaffrey (Research Software Engineer at Microsoft) [61]

Fig. 1. **A researcher's view on the challenges of testing graph algorithm implementations.**

**C1-Vast Input Space.** Consider a straightforward example. The widely used shortest path-finding algorithm, Dijkstra [22], takes a graph, $G$, which can be directed or undirected, edge weights that are non-negative, and an arbitrary source vertex, $s$, as its inputs. The algorithm is supposed to find shortest paths from $s$ to all other vertices on $G$. With $G$ having $M$ edges and $N$ vertices, this renders the input space exceptionally vast to thoroughly test an implementation of the algorithm, surpassing the capacity of human testers. In particular, we would need to select a random graph $G$ and a random vertex $s$ within it.

**C2-Test Oracle Problem.** For many software systems, relying on humans to decide whether they work as expected or not (i.e., using *human oracle*) could be expensive yet reliable [7]. However, when it comes to testing graph algorithm implementations, human testers may struggle to establish ground truth, especially if the input graph is large, owing to its complexity.

Researchers and software developers have long been aware of these challenges, but *(surprisingly?) no solutions have been proposed yet.* Indeed, in 2010, Dr James McCaffrey, a research software engineer at Microsoft, stated, *"Even with a tiny graph, it's easy for a human to make a mistake"* (see Figure 1). In light of recent advancements in fuzzing, which is an effective and efficient automated testing approach [10, 11, 63], we posit these two challenges could be addressed to a certain extent. Notably, code coverage-guided greybox fuzzing (as implemented in popular tools such as AFL [80], AFL++ [28], and LIBFUZZER [12, 38]) has emerged as the state-of-the-art approach in exploring input spaces of large real-world systems (e.g., JavaScript engines, network protocol implementations, embedded systems [24, 40, 71]) leading to the identification of numerous bugs and vulnerabilities. Furthermore, strides have been made in addressing the test oracle problem with recent research showcasing promising outcomes from methodologies such as differential testing and metamorphic testing [18, 70].

Drawing upon these insights, we first develop an AFL-like [80], *code coverage-guided differential fuzzing* tool named *GraphFuzz*$_{\text{cov}}$. We then conduct a preliminary study to assess its effectiveness in testing graph algorithm implementations. *GraphFuzz*$_{\text{cov}}$ discovers six (6) previously unknown bugs, including five (5) logic bugs[2] that require differential testing as the automated test oracle, in the two popular graph libraries NETWORKX and IGRAPH. This result validates *GraphFuzz*_{COV}'s capability to discover previously unknown bugs, which is significant. Nonetheless, the utilization of code coverage as the sole feedback signal encounters two additional challenges.

---

[1]Popular graph libraries such as NETWORKX and IGRAPH support regression testing—which is a form of test automation, not automated testing—and the test cases are constructed manually or are extracted from bug reports.

**C3-Quickly Saturated Code Coverage**. We observe that *GraphFuzz*_{COV} reaches the coverage plateau after just about a few minutes of testing, and the test corpus size is small: (almost) all statements/lines of the algorithm implementations under test can be covered with just a few (normally less than 10) test cases. It suggests that the code coverage feedback signal may not be efficient.

**C4-Multi-language Codebase**. The effectiveness of code coverage-guided fuzzing relies on the correctness and applicability of the coverage-collecting libraries. However, these libraries are normally language specific. For instance, JACOCO [48] only supports Java; GCOV [32] only supports C/C++, and COVERAGE.PY [20] only works with Python. This programming-language dependency makes it hard to collect full coverage data from multi-language implementations like IGRAPH, which is implemented in C/C++, Python, R, and Mathematica.

We address these two additional challenges by extracting and leveraging algorithm-specific feedback signals from the implementations under test. For instance, in fuzz testing strongly connected component algorithms [66], we may consider an input graph $G$ noteworthy and retain it if it produces a new pair $(n, s)$, where $n$ is the number of components and $s$ is the size of the largest component. In Section 4, we provide guidelines for designing such signals and list all the signals defined in this work.

We develop two variants of *GraphFuzz*_{COV} called *GraphFuzz*_{COMBO} and *GraphFuzz*_{ALGO}. In the first variant *GraphFuzz*_{COMBO}, we combine the code coverage feedback with the algorithm-specific feedback and in the second one *GraphFuzz*_{ALGO} we only use the algorithm-specific feedback. While *GraphFuzz*_{COMBO} is designed to address C1, C2, and C3, *GraphFuzz*_{ALGO} could address all the four aforementioned challenges.

Our experiments demonstrate that in the same experimental setup, the two variants *GraphFuzz*_{COMBO} and *GraphFuzz*_{ALGO} can re-discover all six bugs found by *GraphFuzz*_{COV} in a much shorter time. The speedup can be thousands of times faster. We take the best-performing variant, *GraphFuzz*_{ALGO}, to run a longer and broader bug-finding campaign and it discovers 6 more previously unknown bugs.

In summary, this paper makes the following contributions.

- We identify and study four major challenges of automatically testing graph algorithm implementations.
- We design and implement *GraphFuzz*, the first automated framework to test graph algorithm implementations, that address these challenges.
- We conduct large-scale experiments with different settings of *GraphFuzz* and discover 12 previously-unknown bugs in two popular libraries, NETWORKX and IGRAPH.

We make *GraphFuzz* open source at https://github.com/MelbourneFuzzingHub/graphfuzz to facilitate this interesting, yet challenging research direction of automatically testing graph algorithm implementations.

## 2 Background and Example

### 2.1 Graph Definition and Graph Algorithms

A graph is a pair $G = (V, E)$, where $V$ is a finite set of $N$ vertices and $E$ is a set of $M$ edges. If the graph is undirected, each edge is an unordered pair, $(v_1, v_2)$, of vertices $v1, v2$; if the graph is directed, the pair is ordered. In many applications, every edge within a graph is assigned a numerical value. This value could be a cost/weight/distance, in which case, we could solve the shortest-path problem; the value could be a capacity, in which case, we could solve the max-flow/min-cut problem.

---

[2]We define logic bugs as those which, while not causing program crashes, result in incorrect outputs.

There are several libraries implementing graph algorithms. NetworkX [41] and iGraph [21] are widely used general-purpose libraries, among others. Their projects on GitHub have received 14,000+ and 1,700+ stars, respectively. These libraries implement many *graph algorithms* that can be classified into major *categories* such as Path Finding & Search, Community Detection, Centrality & Importance, Similarity, Heuristic Link Prediction and others [64].

**Terminologies (*graph algorithms* vs *graph problems*)** Several *graph algorithms* could be designed and implemented to solve the same *graph problem*. For instance, Dijkstra [22], Bellman-Ford [9], and Goldberg Radzik [33] algorithms all solve the shortest path finding problem, which belongs to the category of Path Finding & Search [64]. Table 1 lists the major algorithms we use to evaluate our framework *GraphFuzz*.

## 2.2 Bugs in Graph Algorithm Implementations

While there is no specific taxonomy of bug types for graph algorithm implementations, yet, our detailed analysis of the 300 most recent issues reported on the GitHub project of the popular graph library NetworkX reveals a multitude of potential pitfalls. Notably, 40% of these issues stem from incorrect logical and functional implementations, leading to unexpected output; 25% are attributed to mishandling of input and output that could lead to program crashes; a further 16% of reported issues pertain to graph generation, layout, and visualization; the remaining 19% result from testing and documentation discrepancies.

**Research Scope.** This paper concentrates on the detection of *crash-triggering bugs* (e.g., due to unhandled exceptions) and *logic bugs*. According to the above analysis, these two types of bugs may account for up to 65% of the reported issues within the NetworkX project. To identify other types of bugs, including security-related bugs and performance bugs, efforts can be directed towards several approaches such as adding assertions, enabling sanitizers like AddressSanitizers [57] and MemorySanitizers [58], or designing new test oracles [10, 11, 60].

As an example, we discuss a previously unknown *logic bug* that was discovered by *GraphFuzz* in the NetworkX's implementation of the Goldberg-Radzik shortest path finding algorithm [34].

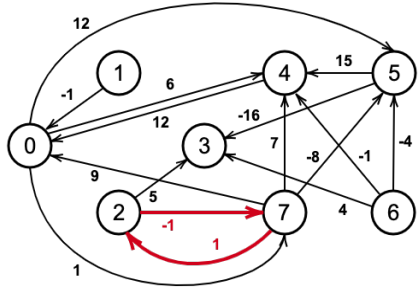

Fig. 2. A logic bug was discovered by *GraphFuzz* in the NetworkX implementation of the Goldberg-Radzik shortest-path finding algorithm [33]. This implementation produces an incorrect output on this graph, *G*, which contains a zero-weight cycle between vertices 2 and 7.

Goldberg-Radzik and Bellman-Ford are two related algorithms that help identify the shortest paths in directed graphs [9, 31, 33]. The Bellman-Ford algorithm allows for negative edges. The

> *"This is puzzling – (so perhaps a bug).* [...] *Manually checking these shows no negative cycles, but the [2,7] cycle has weight 0. I suspect that is the trouble."*
>
> — Prof. Dan Schult (A creator of NetworkX)

Fig. 3. **Prof. Dan Schult's response to our bug report**

Goldberg-Radzik algorithm improves upon the Bellman-Ford algorithm by incorporating techniques for finding the shortest paths in acyclic graphs: if the graph contains a negative cycle, it is found sooner than the Bellman-Ford's algorithm, which does not detect negative cycles until the very end. Therefore, technically, the two algorithms should produce the same results when given the same input: a graph $G$, the source vertex $v$ and target vertex $t$ denoted by $(G, s, t)$.

**A logic bug in the implementation of Goldberg-Radzik's algorithm in NetworkX.** However, *GraphFuzz* successfully generated an input to expose a discrepancy between the implementations of these two algorithms in NetworkX. Given the graph, $G$, shown in Figure 2 and a request to find the shortest path length from (6, 4, 0). Bellman-Ford's implementation successfully returned 11 for the shortest path (6, 4, 0). However, Goldberg-Radzik's implementation reported that there exists a negative cycle. *GraphFuzz* uncovered this bug through the application of the differential testing methodology [62]. It executes both algorithms with identical inputs and identifies any discrepancies between their outputs. Further technical details will be elaborated upon in Section 3 and Section 4.

This bug left the creator and maintainer of NetworkX puzzled (as quoted in Figure 3) because he manually checked the bug-triggering graph that we had reported and found no negative cycles. There are actually six cycles in this graph: (0, 4), (0, 7, 4), (0, 7), (0, 7, 5, 4), (0, 5, 4), (2, 7); all cycles have positive weight, except for the (2, 7) cycle, which has a weight of zero. Indeed, that is the issue because the buggy Goldberg-Radzik's implementation incorrectly checked for negative and zero-weight cycles.

**Why did this bug slip through the testing process?** NetworkX maintains a suite of test cases for regression tests, including test cases for finding the shortest paths on weighted graphs [65]. However, the test suite only covered negative cycles so the tests passed but it failed to test zero cycles. The approved fix reveals that the developer used a less than or equal comparison ($\leq$) when a less than ($<$) comparison was intended. This is not surprising, because developers and testers tend to overlook such corner cases.

## 3 Preliminary Design and Study

### 3.1 *GraphFuzz*$_{\text{cov}}$: Coverage-Guided Differential Fuzzing

Code Coverage-Guided Greybox Fuzzing (CGF) is one of the most successful (security) testing techniques [28, 38, 60, 80]. It is the core technique of the ClusterFuzz framework [36] that has been deployed at large scale to test Google products (e.g., Google Chrome) and popular open-source projects [37]. As of May 2022, ClusterFuzz has found 25,000+ bugs in Google (e.g., Chrome) and 36,000+ bugs in over 550 open source projects integrated with OSS-Fuzz [37]. However, *CGF has not been applied to fuzz test graph algorithm implementations for discovering logic bugs*[3].

To evaluate the effectiveness of CGF in testing graph algorithms, we implement our first version of *GraphFuzz*, called *GraphFuzz*$_{\text{cov}}$, which solely takes code coverage as the feedback. As shown

---

[3]OSS-Fuzz[37] provides CGF fuzzing setups for iGraph and NetworkX but it mainly fuzz tests the parsers (e.g., for input formats like GraphML) of these libraries, not the graph algorithm implementations.

---

**Input:** Algorithm Implementations $A$ & $A'$, Seed Corpus $S$
**Output:** Bug-triggering Graphs $S_x$

1: **repeat**
2:     $g$ = CHOOSE_NEXT($S$)
3:     $p$ = ASSIGN_ENERGY($g$)
4:     **for** $i$ from 1 to $p$ **do**
5:         $g'$ = MUTATE_GRAPH($g$)
6:         $o$ = EXECUTE($A, g'$)
7:         $o'$ = EXECUTE($A', g'$)
8:         **if** $g'$ crashes $A$ or $A'$ **then**
9:             add $g'$ to $S_x$
10:        **else if** IS_DIFFERENT($o, o'$) **then**
11:           add $g'$ to $S_x$
12:        **else if** IS_INTERESTING($A, o, s'$) **then**
13:           add $g'$ to $S$
14: **until** *timeout* reached or *abort*-signal

---

Algorithm 1. The main fuzzing loop of *GraphFuzz*$_{cov}$ and indeed all variants in the *GraphFuzz* framework. It is based on the algorithm modeled by Böhme et al. [14] with changes to support differential testing [62, 70].

in Algorithm 1, *GraphFuzz*$_{cov}$ takes two (and possibly more) algorithm implementations denoted by $A$ and $A'$. It also takes a seed corpus, $S$, which is a set of graphs.

In each iteration of the main fuzzing loop (lines 1–14), *GraphFuzz*$_{cov}$ chooses a graph from the seed corpus (line 2) and decides on the number of new graphs, $p$, to be generated through mutations (line 3). That number is normally called *fuzzing energy* [14]). *GraphFuzz*$_{cov}$ uses different mutation operators to modify the selected graph (line 5). It then executes both $A$ and $A'$ with each newly generated graph and monitors the outputs and behaviors of $A$ and $A'$ to detect or enlarge the seed corpus $S$ (lines 6–13). *GraphFuzz*$_{cov}$ is capable of detecting both crashes and logic bugs (lines 8–11). If $A$ or $A'$ crashes on $g'$, the crash-triggering graph will be saved into $S_x$ for further analysis (lines 8–9). *GraphFuzz*$_{cov}$ detects logic bugs using differential testing [62] (lines 10–11): a bug is triggered if $A$ and $A'$ produce different outputs on the same input. If the newly generated graph $g'$ does not trigger any bug, *GraphFuzz*$_{cov}$ analyzes the coverage result to check if $g'$ can uncover interesting behaviors (e.g., new lines/statements have been executed). It is important to note that, to minimize run-time overhead, *GraphFuzz*$_{cov}$ only checks the coverage result of a single graph implementation, which is $A$ in this case. If so, the graph is retained for further rounds of fuzzing (lines 12–13). Otherwise, *GraphFuzz*$_{cov}$ just discards it. The fuzzing loop ends when a timeout is reached or the user decides to stop the experiment. After that, developers/researchers can analyze the bug-triggering graphs in $S_x$ to identify the root cause and utilize other graphs in the corpus $S$ to calculate code coverage or leverage it for further testing.

### 3.1.1 *Corpus Management*.
Similar to AFL, the seed corpus of *GraphFuzz*$_{cov}$ retains initial inputs as well as those identified as interesting based on the chosen feedback, which are subsequently added during the fuzzing process. It is important to note that inputs could vary across different graph algorithms. For example, a shortest-path finding algorithm may require inputs in the form of $(G, s, t)$[4], whereas a strongly connected component finding algorithm only necessitates a directed graph, $G$.

---

[4]To test single-source multiple-sink/destination shortest path finding algorithms, *GraphFuzz* automatically chooses the sink vertex $t$.

To simplify corpus management in *GraphFuzz*$_\text{cov}$, we only store graphs in the corpus, as they are common input elements across all algorithms. Other input elements, such as $s$ and $t$, are chosen in the test harness in a manner that ensures deterministic results: given the same graph $G$, the algorithm under test—that could take additional information—should produce the same result across runs (we focus here on deterministic outcomes). For instance, in the case of the shortest path finding algorithm, its test harness can select the first vertices that have highest and second-highest degree on $G$ as $s$ and $t$, respectively. By doing so, *GraphFuzz*$_\text{cov}$ can choose the same $s$ and $t$ if the graph $G$ is selected again.

### 3.1.2 *Seed Selection and fuzzing energy assignment*. 
Research has shown that the selection of seeds and the assignment of fuzzing energy can significantly influence the effectiveness and efficiency of coverage/feedback-guided fuzzing, leading to the development of several algorithms aimed at optimizing these processes [12, 14, 54]. However, in this work, we employ a random seed selection algorithm and a fixed, yet configurable fuzzing energy for each selected graph, as our focus lies in evaluating the contribution of feedback signals. A recent study dissecting AFL [30] has highlighted how each design choice impacts various aspects of fuzzing; therefore, we aim to minimize the number of impacting factors. Investigation into the effects of optimized algorithms to *GraphFuzz*$_\text{cov}$ is left for future work.

### 3.1.3 *Mutation Operators*. 
Due to the highly structured nature of graphs, we have devised a set of mutation operators to ensure the validity of the generated graphs. Specifically, we have implemented seven (7) mutation operators, including: (i) adding a vertex, (ii) removing a vertex, (iii) adding an edge, (iv) removing an edge, (v) updating an edge's weight, (vi) trimming the graph, and (vii) combining graphs.

Additionally, we have incorporated the concept of "stacked mutation", following the design principle of the successful fuzzer AFL [30, 80]. Instead of applying a single mutation operator to generate a new graph from a given graph $g$, *GraphFuzz*$_\text{cov}$ randomly selects and applies multiple mutation operators to generate $g'$. The number of operators applied can vary, enabling us to control the level of destructiveness. Sometimes, only one or two operators are selected, resulting in more fine-grained mutations, while at other times, several more operators may be applied, leading to potentially more destructive mutations. In AFL, the number of applied mutations is chosen at random between 2 and 128 [30].

### 3.1.4 *Feedback Signals*. 
At line 12 of Algorithm 1, various feedback signals, potentially including multiple signals, can be utilized to determine the novelty or interest of a newly generated graph. In *GraphFuzz*$_\text{cov}$, we solely rely on code coverage feedback. Specifically, as we implement *GraphFuzz*$_\text{cov}$ in Python, we use statement/line coverage as the default metric within the Coverage.py library [20]. This implies that if the graph $g'$ covers a new statement in the implementation under test, *GraphFuzz*$_\text{cov}$ will save it into the corpus.

### 3.1.5 *Test oracles*. 
In software testing, a test oracle is a procedure used to distinguish between correct and incorrect behaviors of the *System Under Test* [8]. Similar to AFL, *GraphFuzz*$_\text{cov}$ can effectively identify crash-triggering graphs by detecting unhandled exceptions. To detect logic bugs (as exemplified in Section 2), *GraphFuzz*$_\text{cov}$ employs the technique of differential testing [62]. It is important to note that in Algorithm 1, we employ only two algorithm implementations without loss of generality. Technically, *GraphFuzz*$_\text{cov}$ can utilize multiple implementations from the same or different libraries, executing them with generated graphs and identifying any discrepancies. In Section 3.3, we elaborate on our detailed setup, which includes graph implementations from both NETWORKX and IGRAPH.

## 3.2   Implementation

Drawing inspiration from LIBAFL [29], we creat a modular design for *GraphFuzz*$_{cov}$. This design consists of six key components: (i) Corpus, (ii) Scheduler (implementing the CHOOSE_NEXT and ASSIGN_ENERGY functions as outlined in Algorithm 1), (iii) Mutator (implementing MUTATE_GRAPH, (iv) Executor (implementing EXECUTE), (v) Feedback (implementing IS_INTERESTING), and (vi) Tester (implementing test oracles like IS_DIFFERENT).

Like LIBAFL, GraphFuzz is a framework/library to build fuzzers and is not a single general-purpose fuzzer like AFL. To test a new graph problem, a developer can choose suitable components, customize some of them if necessary (e.g., to handle feedback signals differently), and use them as building blocks to create a fuzzer like playing LEGO [27].

We apply the Object-Oriented Programming paradigm to enhance code reusability and reduce the effort needed to implement a fuzzer for new algorithms. Our design includes an abstract BaseFuzzer class that organizes four core components: Corpus, Scheduler, Mutator, and Feedback. To support a new graph problem, a developer only needs to customize the Executor and Tester components to execute the algorithm implementations under test, collect their outcomes to form feedback signals, and detect discrepancies. Since graph problems within the same category share many characteristics, these two components can often be reused with minimal or no modifications.

We implement the whole codebase of the whole *GraphFuzz* framework, including *GraphFuzz*$_{cov}$ and other variants in approximately 3800 lines of Python code.

## 3.3   Experimental Setup

To assess the efficacy of *GraphFuzz*$_{cov}$, we curated a collection of 18 algorithms designed to tackle nine distinct problems spanning six diverse categories. This selection covers five primary categories as listed in the Neo4J graph algorithms infographic [64]. The details are shown in Table 1.

Since *GraphFuzz*$_{cov}$ is a differential fuzzing approach, for each graph problem, we selected at least two graph algorithm implementations. These implementations can be for different algorithms implemented in the same library, or they can come from different libraries, as displayed in the last two columns. For instance, to fuzz test the shortest-path finding problem, we selected two implementations of the Bellman-Ford algorithm: one in NETWORKX and one in IGRAPH. Additionally, we included three other algorithm implementations in NETWORKX: Johnson, Goldberg Radzik, and Floyd Warshall.

We conducted our experiments on an Amazon cloud server running Ubuntu 22.04 64-bit, equipped with 16 virtual CPU cores and 32 GB of memory. Each problem listed in Table 1 underwent 5 trials, each lasting 2 hours, to mitigate the effects of randomness. *GraphFuzz*$_{cov}$ begins with a minimal seed corpus containing only one graph with a single vertex and no edges.

## 3.4   Preliminary Results and Discussions

In this section, we discuss our preliminary results in terms of code coverage achievement and bug detection.

### 3.4.1   *Code Coverage Achievement*. Figure 4 illustrates the mean line coverage achieved by *GraphFuzz*$_{cov}$ within a two-hour timeframe. We notice a consistent trend across all nine problems, where *GraphFuzz*$_{cov}$ reaches a coverage plateau within a few minutes, indicating no further improvement thereafter. Our manual investigation reveals that *GraphFuzz*$_{cov}$ actually covered 95% to 100% of the *reachable* code under test. Surprisingly, this high coverage was attained with a small number of graphs: all corpora have less than 10 graphs. Specifically, the corpus size for SPF, SCC, MFV, MST, JS, MM, HC, AA, and BCC are 5, 3, 9, 3, 3, 3, 4, 3, and 2, respectively.

Table 1. Eighteen graph algorithms curated for accessing *GraphFuzz*. They solve nine graph problems namely Shortest Path Finding (SPF), Minimum Spanning Tree (MST), Strongly Connected Component (SCC), Bi-Connected Component (BCC), Harmonic Centrality (HC), Jaccard Similarity (JS), Max Matching (MM), Adamic Adar (AA), Max Flow Value (MFV).

| Category | Problem | Algorithm | NX | IG |
|---|---|---|---|---|
| Path Finding and Search | Shortest Path Finding | Bellman Ford [9] | ✓ | ✓ |
| | | Goldberg Radzik [33] | ✓ | |
| | | Dijkstra [22] | ✓ | |
| | Minimum Spanning Tree | Prim [73] | ✓ | ✓ |
| | | Kruskal [53] | ✓ | |
| | | Boruvka [16] | ✓ | |
| Community Detection | Strongly Connected Component | Tarjan [77] | ✓ | ✓ |
| | | Tarjan (recursive) | ✓ | |
| | | Kosaraju [76] | ✓ | |
| | Bi-Connected Component | Hopcroft Tarjan [44] | ✓ | ✓ |
| Centrality and Importance | Harmonic Centrality | [15] | ✓ | ✓ |
| Similarity | Jaccard Similarity | [55] | ✓ | ✓ |
| | Max Matching | Hopcroft Karp [45] | ✓ | |
| | | Eppstein | ✓ | |
| | | Push-Relabel [19] | | ✓ |
| Heuristic Link Prediction | Adamic Adar | [55] | ✓ | ✓ |
| Others | Max Flow Value | Goldberg Tarjan [35] | ✓ | ✓ |
| | | Dinitz [23] | ✓ | |
| | | Boykov Komogorov [17] | ✓ | |

> **Finding 1**: The code coverage analysis suggests that relying solely on code coverage as a feedback signal may not be efficient. A small corpus tends to be less diverse, thereby limiting the effectiveness and/or efficiency of bug detection. Additional feedback signals are required to enhance the process.

*3.4.2* ***Bug Detection***. As summarized in Table 5, *GraphFuzz*$_{\text{cov}}$ successfully uncovered six previously unknown bugs, five of which are logic bugs, across five algorithm implementations. Specifically, it detected one logic bug in the Goldberg-Radzik algorithm implementation (coded as Bug-1, explained in Section 2), one logic bug in the Tarjan algorithm implementation for Strongly Connected Components (Bug-2), two bugs (including one logic bug) in the Jaccard Similarity algorithm implementations (Bug-3 and Bug-4), one logic bug in the Max Flow algorithm implementation (Bug-5), and another logic bug in the Adamic-Adar algorithm implementation (Bug-6).

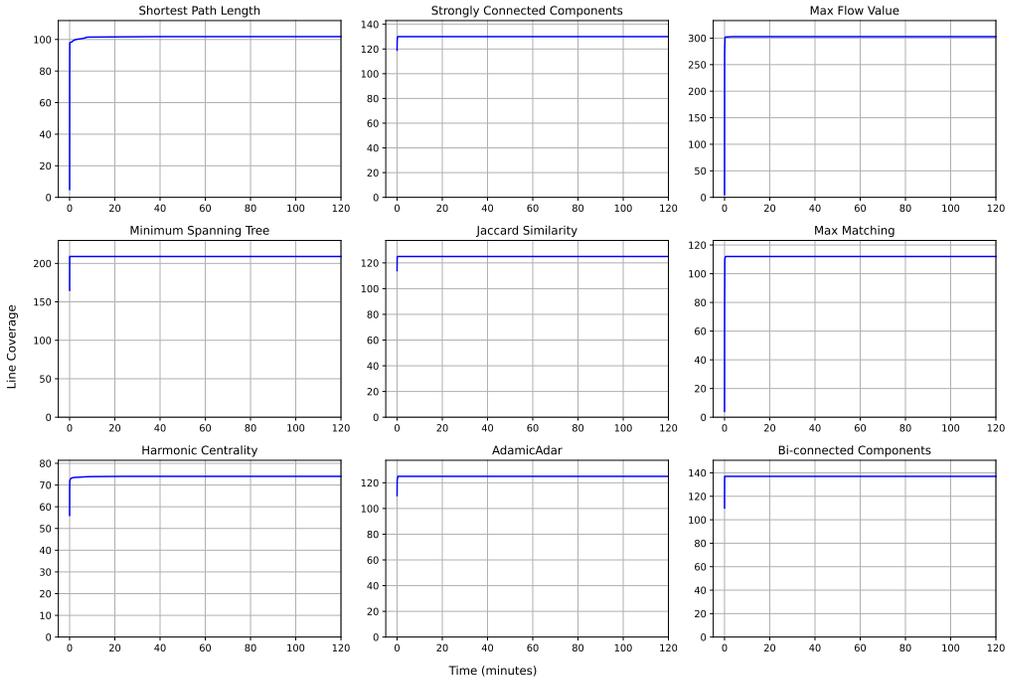

Fig. 4. Mean code coverage over time of nine graph algorithms. We plot the code coverage of one representative graph algorithm implementation in NETWORKX for each graph problem. Specifically we plot the results of Bellman Ford (solving SPF), Prim (MST), Tarjan (SCC), [45] (BCC), [46] (HC), [47] (JS), Hopcroft (MM), [47] (AA), and Goldberg Tarjan (MFV).

Bug-2 affects the computation of strongly connected components, an essential part of analyzing graph structures, as implemented in the Tarjan algorithm (its recursive version in particular). The exact cause of these errors has not been identified yet, making it hard for the NETWORKX team to find a fix. So, the team decided to stop supporting it.

Bug-3 causes the program to crash when running the Jaccard Similarity function with a vertex that does not exist in the graph. This issue happened in both NetworkX and iGraph libraries. Additionally, Bug-4 occurs in the IGRAPH library when using the Jaccard Similarity function on graphs with self-loops. In this case, the function ignores these loops during calculations, leading to wrong results.

Bug-5 happens when the Max Flow Value function is used with a specific input, making the program get stuck in an infinite loop, preventing it from finishing or giving any result. Moreover, there's another logic bug, known as Bug-6, when the Adamic Adar value is calculated on an undirected graph with nodes only connected through self-loops. In these situations, the function gives wrong results for those nodes.

> **Finding 2**: Despite having the limited corpus size due to an inefficient feedback signal, *GraphFuzz*$_{\text{cov}}$ successfully uncovered several previously unknown bugs in popular libraries. This demonstrates the effectiveness of the tool, particularly its mutation operators and its differential testing approach.

## 4 Algorithm-Specific Feedback Signals

Table 2. Feedback signals selected, based on criteria defined in Section 4.

| Graph Problem | Full Output | Feedback | Explanation |
|---|---|---|---|
| SPF | The length of the shortest path between two specified vertices | $(l)$ | $l$: shortest path length |
| SCC | A list of all strongly connected components | $(n, s)$ | $n$: number of components <br> $s$: size of the largest component |
| MFV | The value of the max flow from the source to the sink vertex | $(v)$ | $v$: flow value |
| MST | A minimum spanning tree | $(w, n)$ | $w$: total weight of the MST <br> $n$: number of nodes in the MST |
| JS | A list of scores between all pairs of vertices | $(v)$ | $v$: the highest score |
| MM | A list of matching pairs | $(n)$ | $n$: the number of matching pairs |
| HC | A dictionary of vertices with harmonic centrality as the value | $(d)$ | $d$: the difference between the two smallest scores |
| AA | A list of scores between all pairs of vertices | $(v)$ | $v$: the highest score |
| BCC | A list of all bi-connected components | $(s)$ | $s$: size of the largest bi-connected component |

Based on key insights from Finding 1 in Section 3, we decide to design additional feedback signals to improve the effectiveness and efficiency of our testing methodologies.

To guide the formulation of these signals, we have established several *criteria*. Firstly, the signal should enable us to address challenge C4, as explained in Section 1, thereby facilitating the testing of multi-language libraries or even third-party libraries with no access to source code. Secondly, it should be lightweight to mitigate potential overhead. Thirdly, the signal should distinguish inputs that stress the algorithm in different ways, aiding the testing tool in partitioning the input space into sub-regions. Lastly, but equally importantly, the signal should maintain an appropriate granularity level to ensure the manageability of the test corpus. Recent investigations, as outlined in Herrera et al. [43], discuss the potential drawbacks of overly "fine-grained" signals (e.g., data-flow information like variable names and def-use relationships), which may hinder rather than enhance performance, particularly by overwhelming corpus management and challenge seed selection algorithms.

In alignment with the first and second criteria (being programming-language independent and lightweight), we identify algorithm implementation outputs as potential signals. These outputs can be collected with negligible overhead: collecting them is a part of the testing process. Moreover, implementation outputs can be extracted in a black-box manner which is programming language-independent, and no source code is required.

Additionally, we explore the trade-off between a general signal and algorithm-specific signals. While a general signal may aid in simplifying the implementation of the testing tool, it may not meet the third criterion. Moreover, it is not straightforward to identify a common output produced by all graph algorithms. as exemplified in Table 1, various graph problems yield distinct outputs, such as shortest path lengths versus lists of connected components, underscoring the necessity for tailored signals.

To satisfy the final criterion, and ensure a manageable corpus size, we could opt to select specific parts of the output rather than the entire output or extract some properties of the output, as long as the information still contribute to discriminating the inputs, thereby fulfilling the third criterion. For instance, suppose the output is represented as a triple $(x, y, z)$, where $x$, $y$, and $z$ are 8-bit unsigned integer numbers. In theory, managing a corpus of up to 16,777,216 ($256 \times 256 \times 256$) graphs would be required. However, by selecting only $x$ and $y$, we can reduce this to 65,536 ($256 \times 256$) graphs. Another option is to consider the sum of $x$, $y$, and $z$ as the feedback signal. Moreover, employing further techniques such as the "bucketing" approach in AFL [30, 80], which is used to distinguish loops based on the loop counts, could further diminish the corpus size.

Table 2 lists all algorithm-specific signals we use for the nine graph problems in Table 1. For the Minimum Spanning Tree identification problem, we apply the aforementioned bucketing approach to the total weight of the MST to manage the corpus size more effectively.

In terms of implementation, designing a new feedback signal primarily impacts the Executor component. As explained in Section 3.2, this component is specific to each graph problem: it runs the algorithm implementation under test and extracts information from the output to form signals as outlined in Table 2. Since these signals may share output formats (e.g., an integer for SPF and MFV or a pair of numbers for SCC and MST), we optimize the Feedback component to select interesting test inputs based on data type and format of the signal, making it generic and independent of specific graph problems. Similar to the code coverage signal in *GraphFuzz*$_{\text{COV}}$, a new test input or graph that produces a new signal value is saved in the corpus for further rounds of fuzzing.

## 5   Main Evaluation

### 5.1   Research Questions (with respect to the findings from the preliminary study)

**RQ-1.** Does adding the algorithm-specific feedback signals help improve *GraphFuzz*$_{\text{COV}}$'s performance?

**RQ-2.** How does *GraphFuzz* perform if no feedback signal is in use?

**RQ-3.** How does *GraphFuzz* perform in a larger testing campaign?

In this evaluation, we design three research questions as enumerated above. RQ-1 is designed to evaluate the effectiveness of the proposed algorithm-specific feedback signals. To answer RQ-1, we introduce two new variants of *GraphFuzz*$_{\text{COV}}$ called *GraphFuzz*$_{\text{COMBO}}$, which integrates both the code coverage and algorithm-specific feedback signals, and *GraphFuzz*$_{\text{ALGO}}$, which exclusively relies on algorithm-specific feedback. RQ-2 is devised to assess *GraphFuzz* without any feedback signal, in comparison with other variants of *GraphFuzz*. In this variant called *GraphFuzz*$_{\text{NONE}}$, the fuzzer merely selects graphs from the corpus and mutates them without verifying the novelty of the newly generated graphs, thus not expanding the corpus. This evaluation allows us to ascertain the efficacy of feedback-guided fuzzing in the specific domain of testing graph implementations.

Lastly, we formulate RQ-3 to examine the performance of *GraphFuzz* in terms of bug detection in a larger-scale experiment encompassing more algorithm implementations and a longer duration.

## 5.2 Experimental Setup

To address RQ-1 and RQ-2, we execute the three new variants, *GraphFuzz*<sub>COMBO</sub>, *GraphFuzz*<sub>ALGO</sub>, and *GraphFuzz*<sub>NONE</sub>, on the same set of algorithm implementations under identical settings as evaluated in our preliminary study, as described in Section 3.3, to ensure a fair comparison with *GraphFuzz*<sub>COV</sub>.

Furthermore, we extend the experiments by running all four variants (the three new variants and *GraphFuzz*<sub>COV</sub>) in an additional setup where we began with a larger corpus instead of the small corpus containing only a single-vertex graph used in our preliminary study.

Specifically, for each of the nine graph problems under test, we automatically generate a corpus containing 10 "valid" graphs that adhere to the requirements of the respective graph algorithm implementations. For example, if an algorithm only accepts weighted and directed graphs, we would not include an undirected graph in the initial corpus. This experimental setup allows us to evaluate *GraphFuzz* in a practical context.

In this evaluation, our primary focus lies in evaluating the bug detection capability of *GraphFuzz* across various setups, given its demonstrated ability to rapidly achieve high code coverage across all tested algorithm implementations, as discussed in Section 3.

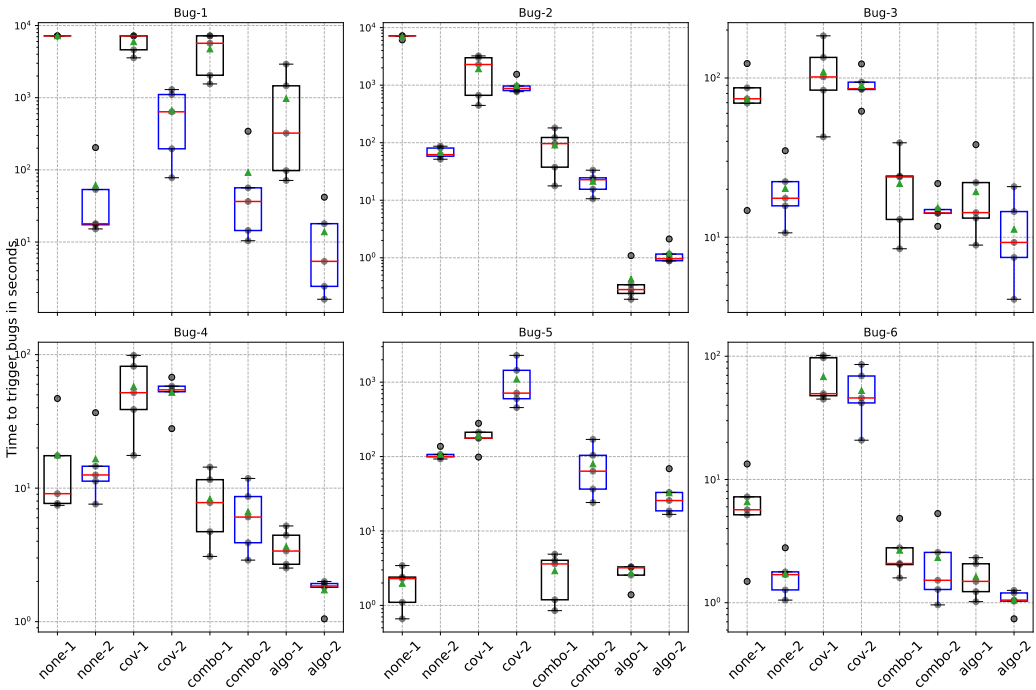

Fig. 5. Boxplots (logarithmic scale) showing the bug finding results of *GraphFuzz* in different setups: none (*GraphFuzz*<sub>NONE</sub>), cov (*GraphFuzz*<sub>COV</sub>), combo (*GraphFuzz*<sub>COMBO</sub>), algo (*GraphFuzz*<sub>ALGO</sub>). Black boxplots (none-1, cov-1, combo-1, algo-1) present the results of *GraphFuzz* with a single-node graph in the initial corpora. Blue boxplots (none-2, cov-2, combo-2, algo-2) present the results of *GraphFuzz* with bigger initial seed corpora.

Table 3. Bug discovery speed-up (based on mean values) of `graphfuzz-none`, `graphfuzz-cov-algo`, and `graphfuzz-algo` over `graphfuzz-cov`. The best result for each bug and corpus is highlighted in bold, while the second-best result is highlighted in italics. Note: "N/A" indicates that the bug was not found within the time limit (2 hours).

|         | **Small Initial Corpus** | | | **Bigger Initial Corpus** | | |
| --- | --- | --- | --- | --- | --- | --- |
|         | **None** | **Combo** | **Algo** | **None** | **Combo** | **Algo** |
| **Bug-1** | N/A | 1.26 | **6.11** | 10.78 | 7.19 | **47.83** |
| **Bug-2** | 0.28 | 21.11 | **4493.07** | 14.62 | 46.54 | **822.10** |
| **Bug-3** | 1.48 | *5.03* | **5.66** | 4.43 | *5.85* | **8.00** |
| **Bug-4** | 3.26 | *6.95* | **15.82** | 3.15 | *7.82* | **30.15** |
| **Bug-5** | **95.36** | 64.65 | *68.46* | *11.63* | 11.29 | **12.40** |
| **Bug-6** | 10.35 | *25.56* | **41.87** | 30.62 | *22.61* | **49.77** |

Table 4. The mean value of fuzzing throughput (i.e., total number of test inputs/graphs generated and evaluated) and corpus size of *GraphFuzz* in different setups, for each 2-hour experiment. *These metrics are measured on the same experiments used to reproduce the six bugs reported in Table 3 and Figure 5.*

|       | **Fuzzing Throughput** | | | | **Corpus Size** | |
| --- | --- | --- | --- | --- | --- | --- |
|       | **None** | **Cov** | **Combo** | **Algo** | **Combo** | **Algo** |
| SPF   | 7579776  | 20540 | 16530 | 44012   | 510   | 1113  |
| SCC   | 19414920 | 25384 | 7223  | 11665   | 186   | 204   |
| MFV   | 9606631  | 20634 | 24884 | 142789  | 717   | 1412  |
| MST   | 13429672 | 12506 | 26922 | 1155797 | 2979  | 10927 |
| JS    | 1879215  | 16678 | 12719 | 55557   | 33    | 42    |
| MM    | 18060231 | 20742 | 60395 | 3442611 | 128   | 154   |
| HC    | 16951825 | 8548  | 8586  | 126522  | 314   | 383   |
| AA    | 18955765 | 14928 | 11229 | 48748   | 483   | 1035  |
| BCC   | 24101405 | 20225 | 20479 | 2961989 | 288   | 294   |

## 5.3 Answers to RQ1

As shown in Table 3 and Figure 5, *GraphFuzz*$_{COMBO}$ and *GraphFuzz*$_{ALGO}$ exhibit the ability to replicate the six bugs discovered by *GraphFuzz*$_{COV}$ in significantly shorter durations. Moreover, the improvement is consistent in both setups of the initial seed corpus, demonstrating the effectiveness of the proposed algorithm-specific feedback signals. When examining the last two columns of Table 4, we observe that the corpus size of the two new variants is much larger yet still manageable compared to the figure reported for *GraphFuzz*$_{COV}$ in Section 3. This increase in corpus size helps improve diversity and could enhance the likelihood of finding more bugs or finding the same bugs faster, as demonstrated in this case.

*GraphFuzz*$_{ALGO}$ outperforms *GraphFuzz*$_{COMBO}$. We investigate the fuzzing throughput of these two setups (as shown in the 4th and 5th columns of Table 4) and find that *GraphFuzz*$_{ALGO}$ generates and evaluates about 40 times more graphs on average than *GraphFuzz*$_{\text{COMBO}}$. This difference could be attributed to the lightweight nature of the proposed feedback signals in comparison to the code coverage feedback. Moreover, the statistics in Table 4 show that the corpus size of *GraphFuzz*$_{\text{ALGO}}$ is, on average, 1.75 times larger than that of *GraphFuzz*$_{\text{COMBO}}$, yet remains manageable, resulting in higher diversity and faster bug discovery.

We notice that having more valid graphs, with respect to the requirements of the graph algorithms, in the initial corpus does help improve the performance of all variants of *GraphFuzz*. It speeds up the discovery for four out of six bugs (as shown in Table 3 and Figure 5). Two exceptions are in finding Bug-2 and Bug-5 in the Strongly Connected Component and Max Flow Value implementation, respectively. We suspect this occurs because these bugs can be triggered with small graphs. When a larger corpus containing bigger graphs is provided, the large graphs tend to be selected and generated more frequently compared to starting with just a single-node graph in the corpus.

> **Finding 3**: The proposed algorithm-specific feedback signals are effective. They help speed up the bug finding process of *GraphFuzz*$_{\text{COV}}$ by up to 4000+ times.

## 5.4 Answers to RQ2: *GraphFuzz*$_{\text{NONE}}$ vs others

The results reported in Table 3 and Figure 5 show that *GraphFuzz*$_{\text{NONE}}$ actually outperforms *GraphFuzz*$_{\text{COV}}$, except in reproducing Bug-1 and Bug-2 with a small initial corpus. As shown in Table 4, its throughput is on average 900 times higher than that of *GraphFuzz*$_{\text{COV}}$ because it does not incur any overhead for either feedback collection or corpus management. With significantly higher throughput and a comparable corpus size, due to the inefficiency of *GraphFuzz*$_{\text{COV}}$ in enhancing the corpus at runtime, the superior performance of *GraphFuzz*$_{\text{NONE}}$ is understandable.

However, *GraphFuzz*$_{\text{NONE}}$ still consistently performs worse than *GraphFuzz*$_{\text{COMBO}}$ and *GraphFuzz*$_{\text{ALGO}}$, despite having throughputs 740 and 19 times higher on average, respectively. This highlights the significance of having a diverse corpus of graphs, as demonstrated by the results of the proposed lightweight program-specific feedback signals.

> **Finding 4**: While *GraphFuzz*$_{\text{NONE}}$—a sole mutation-based fuzzing approach with no feedback guidance—surpasses *GraphFuzz*$_{\text{COV}}$, it falls short of *GraphFuzz*$_{\text{COMBO}}$ and *GraphFuzz*$_{\text{ALGO}}$. While fuzzing throughputs are significant, an effective corpus enhancement strategy, leveraging suitable feedback signals, could prove to be even more crucial.

## 5.5 Answers to RQ3

To address this research question, we selected *GraphFuzz*$_{\text{ALGO}}$—the best performer according to the answers to RQ1 and RQ2—and conducted a set of larger experiments. Specifically, we extended the fuzzing time budget to 24 hours and included several additional algorithm implementations. In addition to the 9 problems listed in Table 1, we tested 17 more graph problems (one implemented in iGraph and 16 implemented in NetworkX) namely Dice Similarity, Panther Similarity, Simrank Similarity, Preferential Attachment, Netative Edge Cycle, All Node Cut, Simple Cycle, Random Spanning Tree, Information Centrality, Intersection Array, Rich Club Coefficient, Effective Size, Global Reaching Centrality, Local Reaching Centrality, Degree Pearson Correlation Coefficient, Random Triad, and Percolation Centrality.

It is worth noting that for the 17 new graph problems, we disabled the differential testing mode for most of them due to the absence of counterpart implementations. For these problems, *GraphFuzz*$_{\text{ALGO}}$ focuses on detecting crashes caused by unhandled exceptions.

After 24 hours, *GraphFuzz*$_{\text{ALGO}}$ detected 18 additional bugs, consisting of three logic bugs and 15 crashes. *GraphFuzz*$_{\text{ALGO}}$ found one more bug in the Goldberg-Radzik algorithm implementation [33] and one more bug in each newly added problem's algorithm implementation. These findings demonstrate the broad applicability and effectiveness of *GraphFuzz* in general, and *GraphFuzz*$_{\text{ALGO}}$ in particular. In the process of reporting these bugs to the maintainers of iGraph and NetworkX, we discovered that while 13 bugs had been previously reported, they remained unresolved or had been addressed in the latest version, which is newer than the version that we tested [5], resulting in 5 confirmed and fixed bugs based on our reports. All bugs and their corresponding links to the bug reports are presented in Table 5 of Appendix A.

Our newly discovered and previously unknown logic bug in the Goldberg-Radzik path finding algorithm implementation is particularly interesting. The public discussion on GitHub[6], suggests this issue exists for several shortest-path functions.

> **Finding 5**: *GraphFuzz*$_{\text{ALGO}}$ is highly applicable and effective. Thus far, we have tested 26 different graph problems including 9 problems reported in Table 1 and 17 problems discussed in this section. These problems cover all five primary categories listed in the Neo4J graph algorithms infographic [64]. Our tool has discovered bugs in 22 of them, resulting in a success rate of 84.6%.

## 6 Threats to validity

The first concern is external validity, specifically the generalizability of our findings. Our results may not apply to graph algorithm implementations that we did not test. However, we evaluated 26 different graph problems, covering all five main categories outlined in the Neo4J graph algorithm infographic [64].

The second concern is internal validity, which refers to how well a study minimizes systematic errors, such as biases in selecting seed corpora or setting up experiments. To address this, we used two types of seed corpora: an empty graph and a larger set of 10 valid graphs. Additionally, we ensured fairness by starting all fuzzers and experimental setups with the same seed corpus and giving each the same time limit. We also repeated each experiment five times to minimize the impact of randomness.

We make our framework open source so that others can test additional algorithm implementations. Furthermore, we share our complete experimental data and detailed instructions for reproducing our results, thereby reducing both of these threats to validity.

## 7 Discussion

### 7.1 Generalizability of algorithm-specific feedback signals

In fuzzing, there are trade-offs between generalizability and effectiveness. For instance, in structure-aware fuzzing [2, 72], having specific grammars or input models significantly improves the results. Another example is metamorphic testing [18]. It requires identifying problem/system-specific metamorphic relations. Compared to the effort of writing grammars or designing MRs, identifying feedback signals in GraphFuzz is much more lightweight. Moreover, developers can use similar feedback signals for graph problems in the same category, saving time and effort.

---

[5]We tested NetworkX version 3.1 and iGraph version 0.11.2
[6]https://github.com/networkx/networkx/issues/7362

We provide a guideline for designing feedback signals in Section 4. Following that guideline, our first author took less than 5 minutes on average to design a suitable feedback signal for each graph problem—not the graph algorithm implementation—under test.

### 7.2 Differential testing vs metamorphic testing

In this work, we have shown that differential testing is effective, though it depends on the assumption that there are at least two similar implementations available. To remove this assumption, applying metamorphic testing [18] is a promising option. For example, we identified a potential metamorphic relation for the shortest-path finding function: given a directed graph $G$ and two vertices $a$ and $b$, if we reverse the directions of all edges in $G$ to obtain $G'$, the shortest-path length from $a$ to $b$ in $G$ should match the shortest-path length from $b$ to $a$ in $G'$. However, extracting metamorphic relations for many other graph algorithm implementations is not straightforward, so we defer exploring this option to future work.

## 8 Related Work

### 8.1 Graph Database Testing

A range of approaches exist to find bugs in graph database systems. While both graph algorithm implementations and graph database systems provide graph-processing capabilities, they differ significantly. Graph database systems provide various expressive graph query languages that allow users to match, analyze, and transform matched subgraphs. Differential testing has been proposed to test the implementation of the Cypher query language [46] as well as for Gremlin [82]. Metamorphic testing has focused on testing graph pattern matching functionality [50, 83] as well as predicates used in queries [52]. Graph frameworks lack query languages, but provide a plethora of graph algorithms geared towards specific tasks, only few of which are implemented by graph database systems. Overall, this makes it challenging to reuse insights of this research strand in our context.

### 8.2 Automated Software Testing with Fuzzing

Since the invention of fuzzing as an automated black-box testing approach in 1990 [63], notable progress has been made to improve the technique, especially in the past 10 years, as evidenced by several significant contributions [10, 60]. These advancements have addressed several challenges, ranging from enhancing its effectiveness and efficiency [12–14] to improving its usability, exemplified by approaches such as automated fuzz driver generation [6, 39, 47, 49, 81]. Our work belongs to another line of important research aimed at broadening the applicability of fuzzing. Recently, several works in this direction have been observed, applying fuzzing effectively to test Deep Learning models [68, 78], Web APIs [4, 67], Database Systems [1], Virtual Devices [56], Embedded Systems [24], JavaScript Engines [40], or even CPUs [75], among others.

### 8.3 Feedback signals for guided fuzzing

The fuzzing research community understands the limitations of relying solely on code coverage as the feedback signal since it can mislead the fuzzer in various cases and overlook interesting inputs [10, 11], as also observed in our preliminary study. Researchers have proposed alternative signals augment the code coverage signals (such as state coverage [71], human-written annotations [3], likely invariants [26], dataflow feedback [43]). They have also proposed signals to replace code coverage when this is difficult or inefficient to collect. For example, in testing database management systems, error codes, query plans can be used as signals [5, 51]. Error code has been used as a feedback signal to test WebGL as well [69]. In this work, we demonstrate that for specific problems like testing graph algorithm implementations, lightweight feedback may be preferable even when code coverage can still be collected. As shown in our evaluation, *GraphFuzz*$_{\text{ALGO}}$ emerges as the top performer.

## 8.4 Test Oracles

We have observed significant progress in the research areas related to designing automated test oracles [8] to detect various types of bugs. Alongside well-known approaches like Metamorphic Testing [18, 42, 67] and differential fuzzing [62, 62], we have also noted other developments. Particularly noteworthy is the creation of new sanitizers that convert silent bugs into detectable ones, such as AddressSanitizer [57], MemorySanitizer [58], and UndefinedBehaviorSanitizer [59], as well as the introduction of new concepts like Intramorphic Testing [74] and Retromorphic Testing [79].

In our work, we utilize differential testing as the test oracle and effectively uncover several logic bugs. In the future, we may explore the application of alternative approaches.

## 9   Conclusion and Future Work

In this study, we have identified and examined the challenges associated with automatically testing graph algorithm implementations. Subsequently, we developed and implemented *GraphFuzz*, the first automated feedback-guided differential fuzzing framework tailored for testing these implementations. Through carefully-designed experiments, we assessed various factors that could influence *GraphFuzz*'s performance, focusing on its feedback mechanism, corpus management, and fuzzing throughput. Our experiments, conducted on two widely used libraries, NETWORKX and IGRAPH, underscore the effectiveness of *GraphFuzz*. The framework successfully discovered 24 bugs, including 8 logic bugs. Of these, 12 bugs were previously unknown and most of them have now been confirmed and fixed by the maintainers.

In our future plans, we intend to expand the application of *GraphFuzz* to test additional graph algorithm implementations. This includes testing within third-party libraries where access to the source code is restricted such as functions handling graph manipulation in database management and navigation systems. With the lightweight feedback signals and versatility of *GraphFuzz*$_{\text{ALGO}}$, our framework could effectively perform these tests, accommodating both black-box and grey-box fuzzing setups.

## References

[1] Peter Alvaro and Manuel Rigger. 2024. Automatically Testing Database Systems: DBMS testing with test oracles, transaction history, and fuzzing. *Queue* 21, 6 (jan 2024), 128–135. https://doi.org/10.1145/3639449

[2] Cornelius Aschermann, Tommaso Frassetto, Thorsten Holz, Patrick Jauernig, Ahmad-Reza Sadeghi, and Daniel Teuchert. 2019. NAUTILUS: Fishing for deep bugs with grammars.. In *NDSS*.

[3] Cornelius Aschermann, Sergej Schumilo, Ali Abbasi, and Thorsten Holz. 2020. Ijon: Exploring deep state spaces via fuzzing. In *2020 IEEE Symposium on Security and Privacy (SP)*. IEEE, 1597–1612.

[4] Vaggelis Atlidakis, Patrice Godefroid, and Marina Polishchuk. 2019. Restler: Stateful rest api fuzzing. In *2019 IEEE/ACM 41st International Conference on Software Engineering (ICSE)*. IEEE, 748–758.

[5] Jinsheng Ba and Manuel Rigger. 2023. Testing database engines via query plan guidance. In *2023 IEEE/ACM 45th International Conference on Software Engineering (ICSE)*. IEEE, 2060–2071.

[6] Domagoj Babić, Stefan Bucur, Yaohui Chen, Franjo Ivančić, Tim King, Markus Kusano, Caroline Lemieux, László Szekeres, and Wei Wang. 2019. Fudge: fuzz driver generation at scale. In *Proceedings of the 2019 27th ACM Joint Meeting on European Software Engineering Conference and Symposium on the Foundations of Software Engineering*. 975–985.

[7] Earl T Barr, Mark Harman, Phil McMinn, Muzammil Shahbaz, and Shin Yoo. 2014. The oracle problem in software testing: A survey. *IEEE transactions on software engineering* 41, 5 (2014), 507–525.

[8] Earl T. Barr, Mark Harman, Phil McMinn, Muzammil Shahbaz, and Shin Yoo. 2015. The Oracle Problem in Software Testing: A Survey. *IEEE Transactions on Software Engineering* 41, 5 (2015), 507–525. https://doi.org/10.1109/TSE.2014.2372785

[9] RICHARD BELLMAN. 1958. ON A ROUTING PROBLEM. *Quart. Appl. Math.* 16, 1 (1958), 87–90. http://www.jstor.org/stable/43634538

[10] Marcel Böhme, Cristian Cadar, and Abhik Roychoudhury. 2020. Fuzzing: Challenges and reflections. *IEEE Software* 38, 3 (2020), 79–86.

[11] Marcel Böhme, Maria Christakis, Rohan Padhye, Kostya Serebryany, Andreas Zeller, and Hasan Ferit Eniser. 2023. *Software Bug Detection: Challenges and Synergies.* Technical Report.

[12] Marcel Böhme, Valentin JM Manès, and Sang Kil Cha. 2020. Boosting fuzzer efficiency: An information theoretic perspective. In *Proceedings of the 28th ACM Joint Meeting on European Software Engineering Conference and Symposium on the Foundations of Software Engineering.* 678–689.

[13] Marcel Böhme, Van-Thuan Pham, Manh-Dung Nguyen, and Abhik Roychoudhury. 2017. Directed greybox fuzzing. In *Proceedings of the 2017 ACM SIGSAC conference on computer and communications security.* 2329–2344.

[14] Marcel Böhme, Van-Thuan Pham, and Abhik Roychoudhury. 2016. Coverage-based greybox fuzzing as markov chain. In *Proceedings of the 2016 ACM SIGSAC Conference on Computer and Communications Security.* 1032–1043.

[15] Paolo Boldi and Sebastiano Vigna. 2014. Axioms for centrality. *Internet Mathematics* 10, 3-4 (2014), 222–262.

[16] Otakar Borůvka. 1926. O jistém problému minimálním. *Práce Moravské Přírodovědecké Společnosti* 3 (1926), 37–58.

[17] Yuri Boykov and Vladimir Kolmogorov. 2004. An experimental comparison of min-cut/max-flow algorithms for energy minimization in vision. *IEEE transactions on pattern analysis and machine intelligence* 26, 9 (2004), 1124–1137.

[18] Tsong Yueh Chen, Fei-Ching Kuo, Huai Liu, Pak-Lok Poon, Dave Towey, TH Tse, and Zhi Quan Zhou. 2018. Metamorphic testing: A review of challenges and opportunities. *ACM Computing Surveys (CSUR)* 51, 1 (2018), 1–27.

[19] Boris V Cherkassky, Andrew V Goldberg, Paul Martin, João C Setubal, and Jorge Stolfi. 1998. Augment or push: a computational study of bipartite matching and unit-capacity flow algorithms. *Journal of Experimental Algorithmics (JEA)* 3 (1998), 8–es.

[20] CoveragePy. 2024. Coverage.py. Retrieved Mar 10, 2024 from https://coverage.readthedocs.io/en/7.4.4/

[21] G Csardi and T Nepusz. 2006. The igraph software. *Complex syst* 1695 (2006), 1–9.

[22] Edsger W Dijkstra. 2022. A note on two problems in connexion with graphs. In *Edsger Wybe Dijkstra: His Life, Work, and Legacy.* 287–290.

[23] Yefim Dinitz. 2006. Dinitz'algorithm: The original version and Even's version. In *Theoretical Computer Science: Essays in Memory of Shimon Even.* Springer, 218–240.

[24] Max Eisele, Daniel Ebert, Christopher Huth, and Andreas Zeller. 2023. Fuzzing Embedded Systems Using Debug Interfaces. In *Proceedings of the 32nd ACM SIGSOFT International Symposium on Software Testing and Analysis.* 1031–1042.

[25] Shimon Even. *Graph algorithms.* Cambridge University Press.

[26] Andrea Fioraldi, Daniele Cono D'Elia, and Davide Balzarotti. 2021. The use of likely invariants as feedback for fuzzers. In *30th USENIX Security Symposium (USENIX Security 21).* 2829–2846.

[27] Andrea Fioraldi and Dominik Maier. 2020. Fuzzers like LEGO. Retrieved Mar 10, 2024 from https://aflplus.plus/rC3_talk_2020.pdf

[28] Andrea Fioraldi, Dominik Maier, Heiko Eißfeldt, and Marc Heuse. 2020. {AFL++}: Combining incremental steps of fuzzing research. In *14th USENIX Workshop on Offensive Technologies (WOOT 20).*

[29] Andrea Fioraldi, Dominik Christian Maier, Dongjia Zhang, and Davide Balzarotti. 2022. Libafl: A framework to build modular and reusable fuzzers. In *Proceedings of the 2022 ACM SIGSAC Conference on Computer and Communications Security.* 1051–1065.

[30] Andrea Fioraldi, Alessandro Mantovani, Dominik Maier, and Davide Balzarotti. 2023. Dissecting American Fuzzy Lop: A FuzzBench Evaluation. *ACM transactions on software engineering and methodology* 32, 2 (2023), 1–26.

[31] Lester Randolph Ford. 1956. Network flow theory. (1956).

[32] GNU. 2024. gcov—a Test Coverage Program. Retrieved Mar 10, 2024 from https://gcc.gnu.org/onlinedocs/gcc/Gcov.html

[33] Andrew V. Goldberg and Tomasz Radzik. 1993. A heuristic improvement of the Bellman-Ford algorithm. *Applied Mathematics Letters* 6, 3 (1993), 3–6. https://doi.org/10.1016/0893-9659(93)90022-F

[34] Andrew V Goldberg and Tomasz Radzik. 1993. *A heuristic improvement of the Bellman-Ford algorithm.* Stanford University, Department of Computer Science.

[35] Andrew V Goldberg and Robert E Tarjan. 1988. A new approach to the maximum-flow problem. *Journal of the ACM (JACM)* 35, 4 (1988), 921–940.

[36] Google. 2024. ClusterFuzz. Retrieved Mar 10, 2024 from https://google.github.io/clusterfuzz/

[37] Google. 2024. Oss-Fuzz. Retrieved Mar 10, 2024 from https://google.github.io/oss-fuzz/

[38] Google Inc. 2015. libFuzzer – a library for coverage-guided fuzz testing. Retrieved October 6, 2023 from https://llvm.org/docs/LibFuzzer.html

[39] Harrison Green and Thanassis Avgerinos. 2022. Graphfuzz: Library API fuzzing with lifetime-aware dataflow graphs. In *Proceedings of the 44th International Conference on Software Engineering.* 1070–1081.

[40] Samuel Groß, Simon Koch, Lukas Bernhard, Thorsten Holz, and Martin Johns. 2023. FUZZILLI: Fuzzing for JavaScript JIT Compiler Vulnerabilities.. In *NDSS*.

[41] Aric Hagberg, Pieter Swart, and Daniel S Chult. 2008. *Exploring network structure, dynamics, and function using NetworkX*. Technical Report. Los Alamos National Lab.(LANL), Los Alamos, NM (United States).

[42] Pinjia He, Clara Meister, and Zhendong Su. 2021. Testing Machine Translation via Referential Transparency. In *Proceedings of the 43rd International Conference on Software Engineering (ICSE '21)*. IEEE Press, Madrid, Spain, 410–422. https://doi.org/10.1109/ICSE43902.2021.00047

[43] Adrian Herrera, Mathias Payer, and Antony L Hosking. 2023. DatAFLow: Toward a data-flow-guided fuzzer. *ACM Transactions on Software Engineering and Methodology* 32, 5 (2023), 1–31.

[44] John Hopcroft and Robert Tarjan. 1973. Algorithm 447: efficient algorithms for graph manipulation. *Commun. ACM* 16, 6 (1973), 372–378.

[45] John E Hopcroft and Richard M Karp. 1973. An n^5/2 algorithm for maximum matchings in bipartite graphs. *SIAM Journal on computing* 2, 4 (1973), 225–231.

[46] Ziyue Hua, Wei Lin, Luyao Ren, Zongyang Li, Lu Zhang, Wenpin Jiao, and Tao Xie. 2023. GDsmith: Detecting bugs in Cypher graph database engines. In *Proceedings of ACM SIGSOFT International Symposium on Software Testing and Analysis (ISSTA)*.

[47] Kyriakos Ispoglou, Daniel Austin, Vishwath Mohan, and Mathias Payer. 2020. {FuzzGen}: Automatic fuzzer generation. In *29th USENIX Security Symposium (USENIX Security 20)*. 2271–2287.

[48] JaCoCo. 2024. JaCoCo Java Code Coverage Library. Retrieved Mar 10, 2024 from https://github.com/jacoco/jacoco/tree/master

[49] Bokdeuk Jeong, Joonun Jang, Hayoon Yi, Jiin Moon, Junsik Kim, Intae Jeon, Taesoo Kim, WooChul Shim, and Yong Ho Hwang. 2023. Utopia: Automatic generation of fuzz driver using unit tests. In *2023 IEEE Symposium on Security and Privacy (SP)*. IEEE, 2676–2692.

[50] Yuancheng Jiang, Jiahao Liu, Jinsheng Ba, Roland H. C. Yap, Zhenkai Liang, and Manuel Rigger. 2024. Detecting Logic Bugs in Graph Database Management Systems via Injective and Surjective Graph Query Transformation. In *Proceedings of the 46th IEEE/ACM International Conference on Software Engineering* (, Lisbon, Portugal,) *(ICSE '24)*. Association for Computing Machinery, New York, NY, USA, Article 46, 12 pages. https://doi.org/10.1145/3597503.3623307

[51] Zu-Ming Jiang, Jia-Ju Bai, and Zhendong Su. 2023. {DynSQL}: Stateful Fuzzing for Database Management Systems with Complex and Valid {SQL} Query Generation. In *32nd USENIX Security Symposium (USENIX Security 23)*. 4949–4965.

[52] Matteo Kamm, Manuel Rigger, Chengyu Zhang, and Zhendong Su. 2023. Testing Graph Database Engines via Query Partitioning. In *Proceedings of the 32nd ACM SIGSOFT International Symposium on Software Testing and Analysis* (, Seattle, WA, USA,) *(ISSTA 2023)*. Association for Computing Machinery, New York, NY, USA, 140–149. https://doi.org/10.1145/3597926.3598044

[53] Joseph B. Kruskal. 1956. On the Shortest Spanning Subtree of a Graph and the Traveling Salesman Problem. *Proc. Amer. Math. Soc.* 7, 1 (1956), 48–50. http://www.jstor.org/stable/2033241

[54] Caroline Lemieux and Koushik Sen. 2018. Fairfuzz: A targeted mutation strategy for increasing greybox fuzz testing coverage. In *Proceedings of the 33rd ACM/IEEE international conference on automated software engineering*. 475–485.

[55] David Liben-Nowell and Jon Kleinberg. 2003. The link prediction problem for social networks. In *Proceedings of the twelfth international conference on Information and knowledge management*. 556–559.

[56] Qiang Liu, Flavio Toffalini, Yajin Zhou, and Mathias Payer. 2023. Videzzo: Dependency-aware virtual device fuzzing. In *2023 IEEE Symposium on Security and Privacy (SP)*. IEEE, 3228–3245.

[57] LLVM. 2024. AddressSanitizer. Retrieved Mar 10, 2024 from https://clang.llvm.org/docs/AddressSanitizer.html

[58] LLVM. 2024. MemorySanitizer. Retrieved Mar 10, 2024 from https://clang.llvm.org/docs/MemorySanitizer.html

[59] LLVM. 2024. UndefinedBehaviorSanitizer. Retrieved Mar 10, 2024 from https://clang.llvm.org/docs/UndefinedBehaviorSanitizer.html

[60] Valentin JM Manès, HyungSeok Han, Choongwoo Han, Sang Kil Cha, Manuel Egele, Edward J Schwartz, and Maverick Woo. 2019. The art, science, and engineering of fuzzing: A survey. *IEEE Transactions on Software Engineering* 47, 11 (2019), 2312–2331.

[61] James D. McCaffrey. 2010. Testing Graph Algorithms. Retrieved Mar 10, 2024 from https://jamesmccaffrey.wordpress.com/2010/01/12/testing-graph-algorithms/

[62] William M McKeeman. 1998. Differential testing for software. *Digital Technical Journal* 10, 1 (1998), 100–107.

[63] Barton P Miller, Lars Fredriksen, and Bryan So. 1990. An empirical study of the reliability of UNIX utilities. *Commun. ACM* 33, 12 (1990), 32–44.

[64] Neo4J. 2024. Graph Algorithms Inforgraphic. Retrieved Mar 10, 2024 from https://go.neo4j.com/rs/710-RRC-335/images/Graph-Algorithms-Infographic.pdf

[65] NetworkX. 2024. Testing shortest path finding on weighted graphs. Retrieved Mar 10, 2024 from https://github.com/networkx

[66] Esko Nuutila and Eljas Soisalon-Soininen. 1994. On finding the strongly connected components in a directed graph. *Information processing letters* 49, 1 (1994), 9–14.

[67] Lianglu Pan, Shaanan Cohney, Toby Murray, and Van-Thuan Pham. 2024. EDEFuzz: A Web API Fuzzer for Excessive Data Exposures. In *Proceedings of the 46th IEEE/ACM International Conference on Software Engineering*. 1–12.

[68] Kexin Pei, Yinzhi Cao, Junfeng Yang, and Suman Jana. 2017. Deepxplore: Automated whitebox testing of deep learning systems. In *proceedings of the 26th Symposium on Operating Systems Principles*. 1–18.

[69] Hui Peng, Zhihao Yao, Ardalan Amiri Sani, Dave Jing Tian, and Mathias Payer. 2023. {GLeeFuzz}: Fuzzing {WebGL} Through Error Message Guided Mutation. In *32nd USENIX Security Symposium (USENIX Security 23)*. 1883–1899.

[70] Theofilos Petsios, Adrian Tang, Salvatore Stolfo, Angelos D. Keromytis, and Suman Jana. 2017. NEZHA: Efficient Domain-Independent Differential Testing. In *2017 IEEE Symposium on Security and Privacy (SP)*. 615–632. https://doi.org/10.1109/SP.2017.27

[71] Van-Thuan Pham, Marcel Böhme, and Abhik Roychoudhury. 2020. Aflnet: a greybox fuzzer for network protocols. In *2020 IEEE 13th International Conference on Software Testing, Validation and Verification (ICST)*. IEEE, 460–465.

[72] Van-Thuan Pham, Marcel Böhme, Andrew E Santosa, Alexandru Răzvan Căciulescu, and Abhik Roychoudhury. 2019. Smart greybox fuzzing. *IEEE Transactions on Software Engineering* 47, 9 (2019), 1980–1997.

[73] R. C. Prim. 1957. Shortest connection networks and some generalizations. *The Bell System Technical Journal* 36, 6 (1957), 1389–1401. https://doi.org/10.1002/j.1538-7305.1957.tb01515.x

[74] Manuel Rigger and Zhendong Su. 2022. Intramorphic testing: A new approach to the test oracle problem. In *Proceedings of the 2022 ACM SIGPLAN International Symposium on New Ideas, New Paradigms, and Reflections on Programming and Software*. 128–136.

[75] Kostya Serebryany, Maxim Lifantsev, Konstantin Shtoyk, Doug Kwan, and Peter Hochschild. 2021. Silifuzz: Fuzzing cpus by proxy. *arXiv preprint arXiv:2110.11519* (2021).

[76] Micha Sharir. 1981. A strong-connectivity algorithm and its applications in data flow analysis. *Computers & Mathematics with Applications* 7, 1 (1981), 67–72.

[77] Robert Tarjan. 1972. Depth-first search and linear graph algorithms. *SIAM journal on computing* 1, 2 (1972), 146–160.

[78] Xiaofei Xie, Lei Ma, Felix Juefei-Xu, Minhui Xue, Hongxu Chen, Yang Liu, Jianjun Zhao, Bo Li, Jianxiong Yin, and Simon See. 2019. Deephunter: a coverage-guided fuzz testing framework for deep neural networks. In *Proceedings of the 28th ACM SIGSOFT international symposium on software testing and analysis*. 146–157.

[79] Boxi Yu, Qiuyang Mang, Qingshuo Guo, and Pinjia He. 2023. Retromorphic Testing: A New Approach to the Test Oracle Problem. *arXiv preprint arXiv:2310.06433* (2023).

[80] Michał Zalewski. 2013. American Fuzzy Lop (AFL). Retrieved October 6, 2023 from https://lcamtuf.coredump.cx/afl/

[81] Mingrui Zhang, Jianzhong Liu, Fuchen Ma, Huafeng Zhang, and Yu Jiang. 2021. Intelligen: Automatic driver synthesis for fuzz testing. In *2021 IEEE/ACM 43rd International Conference on Software Engineering: Software Engineering in Practice (ICSE-SEIP)*. IEEE, 318–327.

[82] Yingying Zheng, Wensheng Dou, Yicheng Wang, Zheng Qin, Lei Tang, Yu Gao, Dong Wang, Wei Wang, and Jun Wei. 2022. Finding Bugs in Gremlin-Based Graph Database Systems via Randomized Differential Testing. In *Proceedings of the 31st ACM SIGSOFT International Symposium on Software Testing and Analysis* (Virtual, South Korea) *(ISSTA 2022)*. Association for Computing Machinery, New York, NY, USA, 302–313. https://doi.org/10.1145/3533767.3534409

[83] Zeyang Zhuang, Penghui Li, Pingchuan Ma, Wei Meng, and Shuai Wang. 2023. Testing Graph Database Systems via Graph-aware Metamorphic Relations. *Proceedings of the VLDB Endowment* 17, 4 (2023), 836–848.

## A    List of Bugs Found by GraphFuzz

Table 5. List of bugs found by GraphFuzz in our experiment. Please note that in our paper, we reported Bug-3 and Bug-7 as the same bug because the root causes are the same. While reporting 13 bugs (Bug-13 to Bug-25) to the maintainers of ɪGRAPH and NETWORKX, we discovered that they had been previously reported but remained unresolved or had already been addressed in a newer version than the one we tested. Specifically, we tested our tool using NETWORKX version 3.1 and ɪGRAPH version 0.11.2. For these bugs, we provide links to the earliest bug reports.

| Index | Algorithm | Logic | Link |
|---|---|---|---|
| **Bug-1** | Shortest Path Finding | ✓ | https://github.com/networkx/networkx/issues/6874 |
| **Bug-2** | Strongly Connected Component | ✓ | https://github.com/networkx/networkx/issues/6897 |
| **Bug-3** | Jaccard Similarity | ✗ | https://github.com/igraph/igraph/issues/2438 |
| **Bug-4** | Jaccard Similarity | ✓ | https://github.com/igraph/igraph/issues/2432 |
| **Bug-5** | Max Flow Value | ✗ | https://github.com/igraph/igraph/issues/2437 |
| **Bug-6** | AdamicAdar | ✓ | https://github.com/igraph/igraph/issues/2448 |
| **Bug-7** | Jaccard Similarity | ✗ | https://github.com/networkx/networkx/issues/7108 |
| **Bug-8** | Shortest Path Finding | ✓ | https://github.com/networkx/networkx/issues/7362 |
| **Bug-9** | Dice Similarity | ✓ | https://github.com/igraph/igraph/issues/2438 |
| **Bug-10** | Panther Similarity | ✗ | https://github.com/networkx/networkx/issues/7108 |
| **Bug-11** | Simrank Similarity | ✗ | https://github.com/networkx/networkx/issues/7108 |
| **Bug-12** | Preferential Attachment | ✗ | https://github.com/networkx/networkx/issues/7108 |
| **Bug-13** | Negative Edge Cycle | ✗ | https://github.com/networkx/networkx/issues/6921 |
| **Bug-14** | All node cut | ✓ | https://github.com/networkx/networkx/issues/6533 |
| **Bug-15** | Simple Cycle | ✓ | https://github.com/networkx/networkx/issues/6052 |
| **Bug-16** | Random Spanning Tree | ✗ | https://github.com/networkx/networkx/issues/6920 |
| **Bug-17** | Information Centrality | ✗ | https://github.com/networkx/networkx/issues/6920 |
| **Bug-18** | Intersection Array | ✗ | https://github.com/networkx/networkx/issues/6920 |
| **Bug-19** | Rich Club Coefficient | ✗ | https://github.com/networkx/networkx/issues/6920 |
| **Bug-20** | Effective Size | ✗ | https://github.com/networkx/networkx/issues/6916 |
| **Bug-21** | Global Reaching Centrality | ✗ | https://github.com/networkx/networkx/issues/6914 |
| **Bug-22** | Local Reaching Centrality | ✗ | https://github.com/networkx/networkx/issues/6914 |
| **Bug-23** | Degree Pearson Correlation | ✗ | https://github.com/networkx/networkx/issues/6913 |
| **Bug-24** | Random Triad | ✗ | https://github.com/networkx/networkx/issues/6915 |
| **Bug-25** | Percolation Centrality | ✗ | https://github.com/networkx/networkx/issues/6886 |